\documentclass[12pt]{article}
\usepackage{a4wide}
\usepackage{amssymb}
\usepackage{graphicx}
\begin{document}
{\renewcommand{\thefootnote}{\fnsymbol{footnote}}
\begin{center}
{\LARGE Information loss, made worse by quantum gravity?}\\
\vspace{1.5em}
Martin Bojowald\footnote{e-mail address: {\tt bojowald@gravity.psu.edu}}
\\
\vspace{0.5em}
Institute for Gravitation and the Cosmos,\\
The Pennsylvania State
University,\\
104 Davey Lab, University Park, PA 16802, USA\\
\vspace{1.5em}
\end{center}
}

\setcounter{footnote}{0}

\begin{abstract}
  Quantum gravity is often expected to solve both the singularity problem and
  the information-loss problem of black holes. This article presents an example
  from loop quantum gravity in which the singularity problem is solved in such
  a way that the information-loss problem is made worse. Quantum effects in
  this scenario, in contrast to previous non-singular models, do not eliminate
  the event horizon and introduce a new Cauchy horizon where determinism
  breaks down. Although infinities are avoided, for all practical purposes the
  core of the black hole plays the role of a naked singularity. Recent
  developments in loop quantum gravity indicate that this aggravated
  information loss problem is likely to be the generic outcome, putting strong
  conceptual pressure on the theory.
\end{abstract}

\section{Introduction}

There is a widespread expectation that quantum gravity, once it is fully
developed and understood, will resolve several important conceptual problems
in our current grasp of the universe. Among the most popular ones of these
problems are the singularity problem and the problem of information
loss. Several proposals have been made to address these questions within the
existing approaches to quantum gravity, but it is difficult to see a general
scenario emerge. Given such a variety of possible but incomplete solution
attempts, commonly celebrated as successes by the followers of the particular
theory employed, it is difficult to use these models in order to discriminate
between the approaches. In this situation it may be more fruitful to discuss
properties of a given approach that stand in the way of resolving one or more
of the big conceptual questions. Here, we provide an example regarding the
information loss problem as seen in loop quantum gravity.

Loop quantum gravity \cite{ThomasRev,Rov,ALRev} is a proposal
for a canonical quantization of space-time geometry. It remains incomplete
because it is not clear that it can give rise to a consistent quantum
space-time picture (owing to the so-called anomaly problem of canonical
quantum gravity). Nevertheless, the framework is promising because it has
several technical advantages compared to other canonical approaches, in
particular in that it provides a well-defined and tractable mathematical
formulation for quantum states of spatial geometry. The dynamics remains
difficult to define and to deal with, but there are indications that a
consistent version may be possible, one that does not violate (but perhaps
deforms) the important classical symmetry of general covariance.  These
indications, found in a variety of models, lead to the most-detailed scenarios
by which one can explore large-curvature regimes in the setting of loop
quantum gravity.

The word ``loop'' in this context refers to the importance attached to closed
spatial curves in the construction of Hilbert spaces for geometry according to
loop quantum gravity \cite{LoopRep}. More precisely, one postulates as basic
operators not the usual curvature components on which classical formulations
of general relativity are based, but ``holonomies'' which describe how
curvature distorts the notion of parallel transport in space-time. If we pick
a vector at one point of a closed loop in curved space and move it along the
loop so that each infinitesimal shift keeps it parallel to itself, it will end
up rotated compared to the initial vector once we complete the loop.  The
initial and final vectors differ from each other by a rotation with an angle
depending on the curvature in the region enclosed by the loop. Loop quantum
gravity extends this construction to space-time and quantizes it: It turns the
rotation matrices into operators on the Hilbert space it provides. An
important consequence is the fact that (unbounded) curvature components are
expressed by {\em bounded} matrix elements of rotations. Most of the
postulated loop resolutions of the singularity problem
\cite{Sing,GenericBounce,QuantumBigBang,BHInt,ModestoConn,HolBH,LoopSchwarz}
rely on this replacement.

Classical gravity, in canonical terms, can be described by a Hamiltonian $H$
that depends on the curvature. If $H$ is to be turned into an operator for
loop quantum gravity, one must replace the curvature components by matrix
elements of holonomies along suitable loops, because only the latter ones have
operator analogs in this framework. One has to modify the classical
Hamiltonian by a new form of quantum corrections. The classical limit can be
preserved because for small curvature, the rotations expressed by holonomies
differ from the identity by a term linear in standard curvature components
\cite{RS:Ham,QSDI}. At low curvature, the classical Hamiltonian can therefore
be obtained. At high curvature, however, strong quantum-geometry effects
result which, by virtue of using bounded holonomies instead of unbounded
curvature, can be beneficial for resolutions of the singularity problem.

Given the boundedness, it is in fact easy to produce singularity-free
models. But one of the outstanding problems of this framework is to show that
the strong modification of the classical Hamiltonian can be consistent with
space-time covariance. This question is not just one of broken classical
symmetries (which might be interesting quantum effects). Covariance is
implemented by a set of gauge transformations which eliminate unphysical
degrees of freedom given by the possibility of choosing arbitrary coordinates
on space-time. When these transformations are broken by quantum effects, the
resulting theory is meaningless because its predictions would depend on which
coordinates one used to compute them. Showing that there are no broken gauge
transformations (or gauge anomalies) is therefore a crucial task regarding the
consistency of the theory. The problem remains unresolved in general, but
several models exist in which one can see how it is possible to achieve
anomaly-freedom, constructed using operator methods
\cite{ThreeDeform,TwoPlusOneDef,TwoPlusOneDef2,AnoFreeWeak,SphSymmOp} or with
effective methods
\cite{ConstraintAlgebra,JR,ScalarHol,ScalarHolInv,HigherSpatial}.

\section{A model of deformed canonical symmetries}
\label{s:Model}

As a simple, yet representative, example, we consider a model with one
field-theoretic degree of freedom $\phi(x)$ and momentum $p(x)$. There is no
room for gauge degrees of freedom in this model, and therefore we use it only
to consider the form of symmetries of gravity, not the way in which spurious
degrees of freedom are removed. 

\subsection{Algebra of transformations}

For the example, we postulate a class of
Hamiltonians
\begin{equation} \label{H}
 H[N]=\int{\rm d}x N\left(f(p)-\frac{1}{4}(\phi')^2-\frac{1}{2}\phi\phi''\right)
\end{equation}
with a function $f$ to be specified, and with the prime denoting a
derivative by the one spatial coordinate $x$. As in general
relativity, the Hamiltonian depends on a free function $N(x)$ because
there is no absolute time. The freedom of choosing $N$ corresponds to
choosing different time lapses and directions along which $H[N]$ would
generate translations. Also the dependence of $H[N]$ on the canonical
fields is modeled on gravity, where $f(p)$ would be a quadratic
function ($p$ standing for extrinsic curvature), and the derivative
terms of $\phi$ present a simple version of spatial curvature (a
function quadratic in first-order and linear in second-order
derivatives of the metric). The main formal features of gravitational
Hamiltonians are therefore captured by this model. One can indeed
check that the general results of
\cite{ConstraintAlgebra,JR,ScalarHol,ScalarHolInv,HigherSpatial}
follow from the structure of derivatives in (\ref{H}) in combination
with a function $f(p)$ which modifies the classical momentum
dependence. 
(Compare with Eq.~\ref{HHbeta} in App.~\ref{a:SphSymm}.)

The Hamiltonian, as a generator of local time translations, is accompanied by
a second generator of local spatial translations, the form of which is more
strictly determined: It is given by $D[w]=\int{\rm d}x w\phi p'$ with another
free function $w(x)$. It generates canonical transformations given by
\begin{equation}
\delta_w\phi=\{\phi,D[w]\}= -(w\phi)' \quad\mbox{and}\quad
\delta_wp=\{p,D[w]\}=-wp'\,,
\end{equation}
as they would result from an infinitesimal spatial shift by $-w(x)$:
\[
p(x-w(x))\approx p(x)-w(x)p'(x)=p(x)+\delta_wp(x)\,.
\]
(The transformation of $\phi$ is slightly different owing to a formal density
weight.)

Of special importance is the algebra of symmetries, which can be computed by
Poisson brackets (as a classical version of commutators). We obtain
\begin{equation} \label{HH}
 \{H[N],H[M]\}= D[{\textstyle\frac{1}{2}}( {\rm d}^2f/{\rm d}p^2)(N'M-NM')]\,.
\end{equation}
Two local time translations have a commutator given by a spatial shift. (The
numerical coefficients chosen in (\ref{H}) ensure that the bracket (\ref{HH})
is closed.) Although our model is simplified, the result (\ref{HH}) matches
well with calculations in models of loop quantum gravity, constructed for
spherical symmetry \cite{JR,HigherSpatial,SphSymmOp} and for cosmological
perturbations \cite{ScalarHol,ScalarHolInv}.
(See Eq.~(\ref{HHbeta}) in App.~\ref{a:SphSymm}.)
The same type of algebra has also been obtained for $H$-operators in
$2+1$-dimensional models \cite{ThreeDeform}. Since our choice
(\ref{H}) extracts the main dynamical features of loop models, it
serves to underline the genericness of deformed symmetry algebras when
$f(p)$ is no longer quadratic.

\subsection{Geometry}

For the classical case in which $f(p)=p^2$ is a quadratic function of $p$,
half the second derivative in (\ref{HH}) is a constant equal to one and the
spatial shift is simply $N'M-NM'$. This relation agrees with the result
obtained in general relativity (except that in the latter case one would have
to use the spatial metric to turn the 1-form $N'M-NM'$ into a vector). It has
an interesting interpretation if we use linear functions of the form $c\Delta
t+ (v/c)x$ for $N$ and $M$ (with the speed of light $c$). The constant $\Delta
t$ amounts to a rigid shift in time. The linear term can be understood if one
thinks of Minkowski diagrams in special relativity: a Lorentz boost tilts the
$x$-axis into a new position by an angle given by the boost velocity $v$. (The
new $x$-axis is the set of points where the new time coordinate
\[
 t'=\frac{t-vx/c^2}{\sqrt{1-v^2/c^2}}
\] 
is constant.) The commutator of Lorentz boosts and time translations can be
derived from (\ref{HH}) with linear $N$ and $M$: For $N=c\Delta t+(v/c)x$ and
$M=-(v/c)x$ (undoing the boost after time $\Delta t$), we have $N'M-NM'=
v\Delta t$. The commutator simply amounts to a spatial shift
\begin{equation} \label{w}
 w=\Delta x=v\Delta t\,,
\end{equation}
as expected.
(It may not be possible to have globally linear functions $N$ and $w$ on a
general manifold, but local Poincar\'e transformations, with $N$ and $w$
linear in some neighborhood, can always be realized.)

Holonomy effects of loop quantum gravity can be modeled by using a bounded
function $f(p)$ instead of a quadratic one. (A popular choice in the field is
$f(p)=p_0^2\sin^2(p/p_0)$ with some constant $p_0$, such as Planck-sized
curvature.) The number of classical symmetries remains intact because the
relation (\ref{HH}) is still a closed commutator. But the structure of
space-time changes: we can no longer think in terms of local Minkowski
geometry because the spatial shift in (\ref{HH}) with $\frac{1}{2}{\rm
  d}^2f/{\rm d}p^2\not=1$ violates the relation $\Delta x=v\Delta t$ found
classically in (\ref{w}). The deviation from classical space-time is
especially dramatic at high curvature, near any maximum of the holonomy
function $f(p)$: Around a maximum, the second derivative is negative, ${\rm
  d}^2f/{\rm d}p^2<0$. For the popular choice of $f(p)=p_0^2\sin^2(p/p_0)$, we
have $\frac{1}{2}{\rm d}^2f/{\rm d}p^2=\cos(2p/p_0)$ which is equal to $-1$ at
the maximum of $f(p)$. The counter-intuitive relation $\Delta x=-v\Delta t$
can be interpreted in more familiar terms: the change of sign means that the
classical Lorentz boost is replaced by an ordinary rotation. (An infinitesimal
rotation by an angle $\theta$ in the $(x,y)$-plane and a spatial shift by
$\Delta y$ commute to $\Delta x=-\theta \Delta y$.) At high curvature,
holonomy-modified models of general relativity replace space-time with pure
and timeless higher-dimensional space, a phenomenon called signature change
\cite{Action,SigChange,PhysicsToday}.

\subsection{Field equations}

At the level of equations of motion, signature change means that hyperbolic
wave equations become elliptic partial differential equations (in all four
dimensions, or two in the model). Indeed, if one computes equations of motion
from the Hamiltonian (\ref{H}), one obtains
\begin{equation} \label{EOM}
 \frac{1}{N} \left(\frac{\dot{\phi}}{N}\right)^{\bullet}-\frac{1}{2}
 \frac{{\rm d}^2f(p)}{{\rm d}p^2} \left(\phi'' +\frac{N'}{N} \phi'+
   \frac{N''}{N}\phi\right)=0\,,  
\end{equation}
where ${\rm d}^2f(p)/{\rm d}p^2$ is a function of $\dot{\phi}$ via
$\dot{\phi}=N{\rm d}f(p)/{\rm d}p$. This partial differential equation, which
is hyperbolic for $\frac{1}{2}{\rm d}^2f(p)/{\rm d}p^2>0$, becomes elliptic
for $\frac{1}{2}{\rm d}^2f(p)/{\rm d}p^2<0$. 

In the latter case, the equation requires boundary values for solutions to be
specified; it is not consistent with the familiar evolution picture
implemented by an initial-value problem. Instead of specifying our field and
its first time derivative at one instant of time, once curvature (or
$\dot{\phi}$ in the model) becomes large enough to trigger signature change we
must specify the field on a boundary enclosing a 4-dimensional region of
interest --- including a ``future'' boundary in the former time direction. We
can no longer determine the whole universe from initial data given at one
time.%
\footnote{A region of signature change can be seen as a barrier to propagation
  and might resemble tunneling in some respects. However, there is a crucial
  difference between these two phenomena: In the barrier of a tunneling
  problem in quantum mechanics, there is a change of sign of a term in the
  relevant partial differential equation, given by the potential minus the
  total energy. This term plays the role of a source term in the partial
  differential equation and does not affect the highest derivative orders. In
  a region of signature change, by contrast, the coefficients of highest
  derivative orders are affected as for instance in (\ref{EOM}). Therefore,
  signature change has important implications for well-posed initial/boundary
  data and causal structures, which are absent in standard tunneling
  problems. Our discussion in the next section relies on these new features.}

Although our specific model is simplified, the main conclusion about signature
change agrees with the more detailed versions cited above, which latter
directly come from reduced models of loop quantum gravity combined with
canonical effective techniques.
(See App.~\ref{a:SphSymm} for an example with spherical symmetry.)
Our model presented here shows that the main reason for signature change is
the modified dependence of gravitational Hamiltonians on curvature components
when holonomies are used to express them, together with the general structure
of curvature terms. (Especially the presence of spatial derivatives seems
crucial for derivatives of the modification function to show up in the
symmetry algebra after integrating by parts.) The rest of our discussions does
not rely on the specific model but rather on the general consequence of
signature change.

\subsection{General aspects of signature change}

As shown in \cite{EffConsQBR}, the structure of constraint algebras or gauge
transformations, of which (\ref{HH}) provides a model, is much less sensitive
to details of regularization effects or quantum corrections than the precise
dynamics implied. Even if there may be additional quantum corrections in
(\ref{EOM}) in a fully quantized model, structure functions of the algebra,
such as $\frac{1}{2}{\rm d}^2f/{\rm d}p^2$ in (\ref{HH}), provide reliable
effects of a general nature. For details, the reader is referred to the above
citation
or App.~\ref{a:EffCons},
but the crucial ingredient in this observation is the definition of
effective constraints $C_I=\langle\hat{C}_I\rangle$ as expectation values of
constraint operators
$\hat{C}_I$ (or symmetry generators $H$ and $D$ in the model here),
and their brackets as $\{C_I,C_J\}=\langle
[\hat{C}_I,\hat{C}_J]\rangle/i\hbar$. A regularization of a constraint
operator $\hat{C}_I$ leads to corresponding modifications of the effective
constraint $\langle\hat{C}_I\rangle$. For any consistent operator algebra, the
bracket of effective constraints mimics the commutator of constraint
operators. Even if $\langle\hat{C}_I\rangle$, computed to some order in
quantum corrections, may give a poor approximation to the quantum dynamics,
the consistent forms of effective constraint algebras restrict the
possible versions of quantum commutators. If effective constraints of a
certain form, such as those obtained with holonomy modifications, always lead
to a change of sign of structure functions, the same must be true for operator
algebras.

As the preceding discussion has made clear, there are no assumptions in this
effective method other than the form of modifications of constraints and that
the theory can be consistent at all. In particular, the phase space or
symplectic structure is not assumed but derived from quantum commutators. The
method therefore provides reliable evaluations of loop models which otherwise
could be analyzed only with difficulties, or in very special circumstances
that provide solvability. Certain proponents of loop quantum gravity often
refer to a ready-made argument in their defense of the theory. If the
effective method shows that there is signature change and corresponding
indeterminacy, they say, then there must be something wrong with this method
\cite{Comm}. It is therefore important to realize that the effective method is
merely used to evaluate loop models. It does make assumptions, but only of
general type: it assumes that it is possible to have some anomaly-free
realization of constraint algebras, and that sufficiently general
semiclassical states exist which allow one to derive mode equations with
quantum corrections. (The latter assumption is necessary in
background-independent theories, but not only for their effective analysis.)
It is also important that the effective method is the only one so far that has
provided results on the off-shell constraint algebra of loop quantum
gravity. The constructions by \cite{TwoPlusOneDef,TwoPlusOneDef2,AnoFreeWeak}
are only partially off-shell at isolated points, in such a way that they do
not show deformations of the constraint algebra by holonomy corrections, which
are responsible for signature change. The constraint analysis of black-hole
models by \cite{LoopSchwarz} makes use of a simple Abelianization which is
available only in this specific situation, and only for the classical
constraints. Although these models implement holonomy modifications after the
classical constraints have been Abelianized, they do not allow conclusions
about quantum space-time as given by holonomy-modified constraints.

As noted also in \cite{SigChange,DeformedCosmo}, equations of the form
(\ref{EOM}) sometimes appear for matter systems with instabilities, in
cosmology but also in other areas such as transonic flow. An instability would
normally not be interpreted as signature change as long as a standard
Lorentzian metric structure remains realized, as is the case in all the known
matter examples. The present context, however, is different, because the
instability affects the geometry of space-time itself, and not just matter
propagating in space-time. (In models of loop quantum gravity, $\phi$ in
(\ref{EOM}) stands for metric inhomogeneities.) Such an instability is more
severe, and at the same time more inclusive because it affects all excitations
--- matter and geometry --- in the same way. Indeed, the most fundamental
structure where it appears is not the equation of motion (\ref{EOM}) but the
symmetry algebra (\ref{HH}). If matter is present, its Hamiltonian would be
added to the gravitational one, the resulting sum satisfying a closed algebra
of the form (\ref{HH}). (If adding matter terms would break the algebra, there
would be anomalies making the theory inconsistent.) Matter and geometry are
then subject to the same modified symmetries, and correspondingly to a
modified evolution picture with a boundary rather than initial-value problem
at high density.

Solutions might exist for elliptic partial differential equations with an
initial-value problem. However, such solutions are unstable and depend
sensitively on the initial values; therefore, initial-value problems for
elliptic partial differential equations are not well-posed. Sometimes, a
physical model of this form may just signal a growing mode which is increasing
rapidly in actual time. In quantum gravity and cosmology, however,
instabilities from signature change in (\ref{HH}) or (\ref{EOM}) are much more
debilitating. In this context, one does not perform controlled laboratory
experiments in which one can prepare or directly observe the initial
values. When signature change is relevant, it happens in strong
quantum-gravity regimes where the analogs of $f(p)$ differ much from the
classical behavior. Not only initial values but also the precise dynamical
equations (subject to quantization ambiguities) are so uncertain that an
initial-value formulation can give no predictivity. (In cosmological parlance,
instabilities from signature change present severe versions of trans-Planckian
and fine-tuning problems. For more information on the dynamics of affected
modes see \cite{FluctEn}.) In contrast to some matter systems in which
elliptic field equations may appear, quantum-gravity theories do not allow
initial-value formulations in such regimes but rather require 4-dimensional
boundary-value problems.

Evolution in these models is no longer fully deterministic. In the remainder of
this article, we apply this conclusion to black holes and show that even
low-curvature regions, where observers have no reason to expect strong
quantum-gravity effects, will be affected by indeterminism.  In this context,
consequences of signature change are therefore much more severe than their
analogs in cosmological models.

\section{Black holes}

Black holes in general relativity have singularities where space-time
curvature diverges. Loop quantum gravity has given rise to models in which
curvature is bounded, apparently resolving the singularity problem
\cite{BHPara}. As in some other approaches
\cite{ClosedHorHighDer,ClosedHor,ClosedHorTrapping,InfoLossVaidya}, there is
then no event horizon but only an apparent horizon which encloses large
curvature but eventually shrinks and disappears. If there is no singularity
and information can travel freely through high-curvature regions, there is no
information loss, so this important problem seems to be resolved too. However,
previous black-hole models of this type in loop quantum gravity did not
consider the anomaly problem. In an anomaly-free version, curvature may still
be bounded, but when it is large (Planckian, or near the upper bound provided
by the models), there can be signature change, preventing information from
travelling freely through this regime. It is no longer obvious that the
information-loss problem can be resolved in singularity-free models of black
holes.

\begin{figure}
\begin{center}
\includegraphics[width=4.5cm]{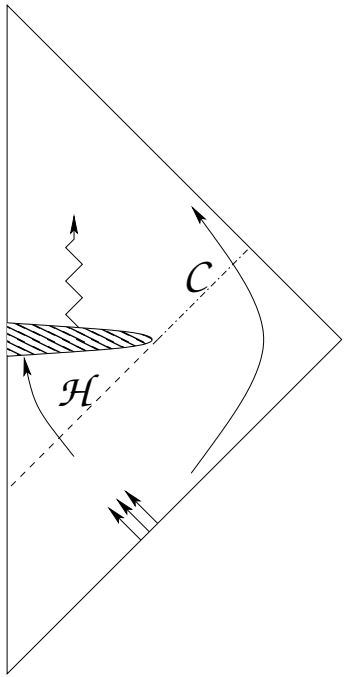}
\caption{Acausality: Penrose diagram of a black hole with signature change at
  high 
  curvature (hashed region). In contrast to traditional non-singular models,
  there is an event horizon (dashed line ${\cal H}$, the boundary of the
  region that is determined by backward evolution from future infinity)
  and a Chauchy horizon (dash-dotted line ${\cal C}$, the boundary of the
  region obtained by forward evolution of the high-curvature region). After an
  observer crosses the Cauchy horizon, space-time depends on the data
  chosen on the top boundary of the high-curvature region and is no longer
  determined completely by data at past infinity. Information that falls
  through ${\cal H}$ affects field values in the hashed region, but not on the
  top boundary or its future; it is therefore lost for an outside
  observer. Unrestricted boundary values at the top part of the hashed region
  influence the future universe even at low curvature (zigzag arrow), a
  violation of deterministic behavior.
  \label{Fig:Cauchy} }
\end{center}
\end{figure}

If the singularity is resolved, there are two scenarios for
Hawking-evaporating black holes: The black-hole region enclosed by an apparent
horizon could reconnect with the former exterior at the future end of high
curvature, or it could split off into a causally disconnected baby
universe. The latter case does not solve the information-loss problem because
information that falls into the black hole is sealed off in the baby
universe. The former case resolves the information-loss problem only if
information can travel through high curvature. 
(In this article, we leave aside other problems of such remnant scenarios,
as discussed for instance in \cite{Giddings}, and focus on issues that
originate from the same mechanism that is used to remove singularities.)
If signature change happens,
nothing travels through the high-curvature region and the fate of information
must be reconsidered.

The elliptic nature of field equations in the high-curvature core of black
holes requires one to specify fields at the future boundary, which would
evolve into the future space-time after black-hole evaporation. 
(The signature-change models analyzed here may also arise as effective
versions of wave-function collapse models proposed in \cite{Coll1,Coll2}. Free
boundary data around the high-curvature core would then correspond to the
undetermined wave function obtained by quantum collapse.)
In Fig.~\ref{Fig:Cauchy}, boundary values on the bottom line surrounding the
hashed high-curvature region would be determined by evolving past initial
values forward in time, but boundary data on the top line around the region
would have to be specified, unrestricted by any field equations. Their values
are not predicted by the theory, and yet they are essential for determining
the future space-time. Once the high-curvature region is passed by an outside
observer, space-time is no longer predictable. The black-hole's event horizon
${\cal H}$ extends into a Cauchy horizon ${\cal C}$: The region above ${\cal
  C}$ is affected by undetermined boundary data. Even if there are no
infinities, the classical black-hole singularity is, for practical purposes,
replaced by a naked singularity, a place out of which unpredictable fields can
emerge.

In terms of information loss, whatever infalling matter hits the
high-curvature core of the black hole determines some part of the boundary
conditions required for the elliptic region, and thereby influences part of
the solution in the core. But it does not restrict our choice for the future
boundary data, or anything that evolves out of it at lower
curvature. Infalling information is therefore lost even if there is no
black-hole singularity
in the sense of infinite curvature.
Similar conclusions apply to the alternative of a baby universe:
Infalling information cannot be retrieved in the old exterior, and it
cannot be passed on to the baby universe.

\section{Conclusions: A no-heir theorem?}

We have presented here a mechanism which appears to be generic in loop quantum
gravity and helps to resolve curvature divergence, but makes the information
loss problem of black holes worse. 
(Interestingly, models of string theory have occasionally resulted in similar
effects of signature change or related phenomena
\cite{SigSingString,Cheshire}.)
Black-hole singularities can turn into naked singularities in this framework,
which implies an end to predictivity. In classical general relativity, there
is strong evidence that cosmic censorship applies: given generic initial data,
singularities may form but are enclosed by black-hole horizons; no naked
singularities appear that would affect observations made from far away. In
loop quantum gravity, a stronger version of cosmic censorship would be
required if signature change is confirmed to be generic. Naked singularities
(Cauchy horizons) could be avoided only if black-hole interiors split off into
baby universes. But even then, information could not be passed on to the baby
universe. From the point of view of observers in this new universe, the former
black-hole singularity would appear as a true beginning, just as the big bang
appears to us in our universe.

The information-loss problem has turned into a more-severe problem of
indeterminism. Two options remain for loop quantum gravity to provide a
consistent deterministic theory without Cauchy horizons. First, one might be
able to show that signature change does not happen under general conditions in
the full theory
or that non-perturbative effects in $\hbar$ somehow allow for deterministic
propagation,
a question which requires an understanding of the off-shell
constraint algebra and the thorny anomaly problem,
together with the equally difficult problem of non-perturbative physical
observables.
All current indications, however, point in the opposite direction and suggest
that signature change is generic. With signature change, Cauchy horizons can
be avoided only if the high-curvature regions of black holes always remain
causally disconnected from the universe in which they formed, that is if black
holes open up into new baby universes. In this scenario, information that
falls in a black hole is still lost even for the baby universe, but at least
the more-severe problem of a Cauchy horizon can be avoided. In either case, a
detailed analysis of possible consistent versions of the constraint algebra of
loop quantum gravity could lead to a ``no-heir theorem'' if deterministic
evolution through the high-density regime of black holes turns out to be
impossible under all circumstances. Black holes would have no heirs since
everything possessed by a collapsing star, including the information carried
along, would be lost even if space-time did not end in a curvature
singularity.

So far, loop quantum gravity is not understood sufficiently well for a clear
model of black holes to emerge from it, but the mechanism analyzed here shows
that, at the very least, scenarios obtained from generalizations of simple
homogeneous models, such as the one postulated in \cite{BHPara}, are likely to
be misleading. Inhomogeneity can change the picture drastically, not just
because there may be back-reaction on a homogeneous background but
also, and often more surprisingly, because the non-trivial nature of symmetry
algebras such as (\ref{HH})
or (\ref{HHbeta})
is much more restrictive for inhomogeneous models. (The right-hand
side would just be identically zero with homogeneity, hiding the
crucial coefficient and its sign which indicates signature change.)
Our considerations of black-hole models provide a concrete physical
setting in which loop quantum gravity and its abstract anomaly problem
can be put to a clear conceptual test.

\section*{Acknowledgements}

This work was supported in part by NSF grant PHY-1307408. The author thanks
Steve Giddings, Karl-Georg Schlesinger and Daniel Sudarsky for interesting
suggestions about related work.

\begin{appendix}

\section{Anomaly-free constraint algebras in spherically symmetric models}
\label{a:SphSymm}

This appendix recalls further details of anomaly-free constraints derived for
spherically symmetric models of gravity. The details of variables and
constraints are more contrived than in the toy model of Section~\ref{s:Model},
but the same key features are realized.

Spherically symmetric triad variables and their momenta can be parameterized
by two pairs of fields $(K_x,E^x)$ and $(K_{\varphi},E^{\varphi})$ depending
on a radial coordinate $x$, with $E^x$ and $E^{\varphi}$ components of a
densitized triad and $K_x$ and $K_{\varphi}$ parameters for extrinsic
curvature \cite{SphSymm}.  The Hamiltonian constraint with potential
modifications from holonomy effects of loop quantum gravity can be
parameterized as \cite{JR,LTBII,HigherSpatial}
\begin{eqnarray} \label{HQsph}
H[N]&=&-\frac{1}{2G}\int {\rm d} x\,
N\bigg(|E^x|^{-\frac{1}{2}} E^{\varphi}f_1(K_{\varphi},K_x)+
2|E^x|^{\frac{1}{2}}f_2(K_{\varphi},K_x) \nonumber\\ 
&&\qquad +|E^x|^{-\frac{1}{2}}(1-\Gamma_{\varphi}^2)E^{\varphi}+
2\Gamma_{\varphi}'|E^x|^\frac{1}{2} \bigg)
\end{eqnarray}
with the spin-connection component
$\Gamma_{\varphi}=-(E^x)'/2E^{\varphi}$ and two functions $f_1$ and
$f_2$ of extrinsic curvature (or possibly the triad as well).

Classically, $f_1=K_{\varphi}^2$ and $f_2=K_xK_{\varphi}$, but this is not the
only possibility of anomaly-free constraint algebras, together with the
diffeomorphism constraint
\begin{equation} \label{Dsph}
 D[N^x]=\frac{1}{2G} \int{\rm d}x N^x
 \left(2E^{\varphi}K_{\varphi}'-K_x(E^x)'\right)\,.
\end{equation}
It remains unknown how the linear dependence on $K_x$ of the classical
constraint can be modified in an anomaly-free way (which, as shown in
\cite{HigherSpatial}, would likely involve higher spatial
derivatives). But the dependence on $K_{\varphi}$ is not uniquely determined
by anomaly-freedom alone.  If $f_2=K_x F_2(K_{\varphi},E^x,E^{\varphi})$ with
$F_2$ related to the free function $f_1$ by \cite{JR}
\begin{equation} \label{alphaGamma}
F_2+ 2E^x \frac{\partial F_2}{\partial E^x}=
\frac{1}{2}\frac{\partial f_1}{\partial K_{\varphi}}\,,
\end{equation}
there is a closed algebra
\begin{equation} \label{HHbeta}
\{H[N],H[M]\} = D[\beta h^{ab}(N_1\partial_bN_2-N_2\partial_bN_1)]
\end{equation}
with
\begin{equation} \label{betasph}
 \beta=\frac{\partial
F_2}{\partial K_{\varphi}}
\end{equation}
a function of the canonical variables.  In the simple case in which $F_2$ does
not depend on $E^x$, we have
\begin{equation} \label{betasph2}
\beta(K_{\varphi})= \frac{1}{2}
\frac{\partial^2 f_1}{\partial K_{\varphi}^2}
\end{equation}
of the form (\ref{HH}). 

The same relations of modification functions and deformed algebras can
be derived at the operator level \cite{SphSymmOp}. Therefore,
consequences such as signature change are not restricted to effective
derivations, but they are much easier to see in this way based on the
mode equations implied. (More generally, signature change and other
properties of quantum space-time structure are reflected abstractly in
the structure functions of hypersurface-deformation algebras such as
(\ref{HHbeta}).)  Similar results have been derived for linear
perturbations in cosmological models \cite{ScalarHolInv}; see
\cite{DeformedCosmo} for a detailed comparison with spherically
symmetric models.

Field equations generated by a modified constraint (\ref{HQsph}) (with
(\ref{Dsph})) could be used to find explicit solutions for the model
pictured in Fig.~\ref{Fig:Cauchy}. However, details of solutions
depend on a large set of quantization ambiguities summarized here in
the free function $f_1(K_{\varphi})$. They would also be sensitive to
additional corrections from quantum back-reaction of fluctuations and
higher moments. The qualitative model of Fig.~\ref{Fig:Cauchy}, on the
other hand, is robust: for any bounded $f_1(K_{\varphi})$, as implied
by holonomy modifications, $\beta$ is negative near a local maximum of
curvature according to (\ref{betasph2}). Moreover, as shown in
\cite{EffConsQBR} and App.~\ref{a:EffCons}, $\beta$ is not subject to
quantum back-reaction from moments.

\section{Canonical effective theory and constraints}
\label{a:EffCons}

Canonical effective methods \cite{EffAc} evaluate dynamical equations
for expectation values and moments of a state based on algebraic
properties of quantum operators. No specific Hilbert-space
representation is assumed, implying key advantages especially for
constrained systems where physical Hilbert spaces are often difficult
to derive.

Starting with a $*$-algebra ${\cal A}$ generated by some basic operators
$\hat{A}_i$, $i=1,\ldots,n$, a state is a positive linear functional
$\langle\cdot\rangle\colon{\cal A}\to{\mathbb C}$. That is,
$\langle\hat{A}^*\hat{A}\rangle\geq 0$ for all $\hat{A}\in{\cal A}$. (The
$*$-relation is abstractly defined, here for simplicity assuming
$\hat{A}_i^*=\hat{A}_i$ for self-adjoint generators. In a Hilbert-space
representation, the $*$-relation would be given by taking adjoint operators.) 

Instead of working with entire states, we can paramaterize them by an infinite
set of numbers given by the expectation values $\langle\hat{A}_i\rangle$ of
basic operators and the moments
\begin{equation} \label{Moments}
 \Delta(A_1^{a_1}\cdots A_n^{a_n}) =
 \langle(\hat{A}_1-\langle\hat{A}_1\rangle)^{a_1}\cdots
 (\hat{A}_n-\langle\hat{A}_n\rangle)^{a_n}\rangle_{\rm symm}
\end{equation}
with products taken in totally symmetric ordering so as to remove
redundancies. (For $a_1+\cdots+a_n=2$, the resulting moments are fluctuations
and covariances.)

The set of basic expectation values and moments is turned into a quantum phase
space by introducing a Poisson bracket by
\begin{equation} \label{Poisson}
 \{\langle\hat{A}\rangle,\langle\hat{B}\rangle\} := \frac{\langle
   [\hat{A},\hat{B}]\rangle}{i\hbar}
\end{equation}
for expectation values and extending it to products as in (\ref{Moments})
using the Leibniz rule.

The expectation value of a combination of the basic operators (in totally
symmetric ordering), such as a constraint $\hat{C}_I=C_I(\hat{A}_i)$, can be
written as a function on the quantum phase space by a formal Taylor expansion
\begin{eqnarray}
 \langle\hat{C}_I\rangle &=& \langle C_I(\hat{A}_i)\rangle= \langle
 C_I(\langle\hat{A}_i\rangle+
 (\hat{A}_i-\langle\hat{A}_i\rangle))\rangle\nonumber \\
&=& C_I(\langle\hat{A}_i\rangle)+ \sum_{a_1,\ldots,a_n} \frac{1}{a_1!\cdots a_n!}
\frac{\partial^{a_1+\cdots+a_n} C_I(\langle\hat{A}_i\rangle)}{\partial
  \langle\hat{A}_1\rangle^{a_1}\cdots \partial\langle \hat{A}_n\rangle^{a_n}}
\Delta(A_1^{a_1}\cdots A_n^{a_n})\,. \label{Expand}
\end{eqnarray}
(For polynomial $C_I$, the sum is finite and an exact representation of the
left-hand side.)

Seen as functions of basic expectation values and moments as per
(\ref{Expand}), the condition that $\langle\hat{C}_I\rangle=0$ in a
physical state ($\hat{C}_I\psi=0$ in a Hilbert-space representation)
provides a constraint function on the quantum phase space
\cite{EffCons,EffConsRel}. Moreover, any expression of the form
$\langle\hat{p}\hat{C}_I\rangle=0$ must vanish in a physical state but
generically provides a constraint independent of
$\langle\hat{C}_I\rangle=0$ if $p\not=1$. Every constraint operator
$\hat{C}_I$ generates an infinite set of independent constraints on
the quantum phase space, which can be organized by the polynomial
degree of $\hat{p}$ restricted to polynomials in basic operators
$\hat{A}_i$. (One can easily see that an infinite set of constraints
is necessary in order to restrict not just the basic expectation value
$\hat{A}_i$ but also all the $A_i$-moments if the classical $A_i$ is
totally constrained by $C_I$.)

By definition of the Poisson bracket (\ref{Poisson}) on the quantum phase
space, the quantum commutators of constraints $\hat{C}_I$, and therefore the
whole constraint algebra, are faithfully mapped to corresponding
Poisson-bracket relations of effective constraints obtained by applying
(\ref{Expand}). It is usually easier to compute Poisson brackets than
commutators, especially in the presence of ordering and other
ambiguities. Moreover, if one truncates (\ref{Expand}) to finite orders of
moments up to some maximum order, one can obtain order-by-order information on
the constraint algebra in a semiclassical expansion. (For a semiclassical
state, a moment of order $a_1+\cdots+ a_n$ behaves as
$O(\hbar^{(a_1+\cdots+a_n)/2})$, as can be checked explicitly for a Gaussian.)
Some key properties of constraint algebras are independent of the order, and
therefore provide information suitable for strong quantum regimes. In this
paper, the main result of \cite{EffConsQBR} is important, which states that
structure functions in constraint algebras of the form
$\{C_I,C_J\}=f^K_{IJ}(A_i)C_K$ with phase-space dependent $f_{IJ}^K$ (such as
the inverse metric $h^{ab}$ in (\ref{HHbeta})) do not receive quantum
corrections by moments.

As a brief justification of this result, one may note that the effective
algebra reads
\begin{equation}
 \{\langle\hat{C}_I\rangle,\langle\hat{C}_J\rangle\} =
 \langle\hat{f}_{IJ}^K\hat{C}_K\rangle
\end{equation}
if $\hat{f}_{IJ}^K$ quantize the classical structure functions. One
could expect moment-dependent corrections of the structure functions
if the right-hand side could be written with a factor of
$\langle\hat{f}_{IJ}^K\rangle$ and $f$ is non-linear in
$A_i$. However, expanding as in (\ref{Expand}) in such a way that all
resulting terms are proportional to effective constraints
$\langle\hat{p}\hat{C}_I\rangle$ (but not
$C_I(\langle\hat{A}_i\rangle)$ which need not vanish for physical
states), we have
\begin{equation} \label{fC} 
\{\langle\hat{C}_I\rangle,\langle\hat{C}_J\rangle\}=
  f_{IJ}^K(\langle\hat{A}_i\rangle) \langle\hat{C}_K\rangle
  + \sum_j \frac{\partial
    f_{IJ}^K(\langle\hat{A}_i\rangle)}{
 \partial\langle\hat{A}_j\rangle} \langle\hat{A}_j\hat{C}_K\rangle+\cdots
\end{equation}
with higher-order constraints $\langle\hat{A}_j\hat{C}_K\rangle$ and
so on, but no moment corrections in the coefficients
$f_{IJ}^K(\langle\hat{A}_i\rangle) $ of
$\langle\hat{C}_K\rangle$. Also the higher-order constraints which
appear due to the additional quantum degrees of freedom have
coefficients such as $\partial f_{IJ}^K(\langle\hat{A}_i\rangle)/
\partial\langle\hat{A}_j\rangle$ independent of moments. In other
words, quantum back-reaction is realized in effective constraint
algebras not by moment corrections in structure functions, but by an
extension of the algebra to quantum degrees of freedom. In spherically
symmetric models, $\beta$ in (\ref{betasph2}) is not modified by
quantum back-reaction and holds to all orders in an
$\hbar$-expansion. Effects which depend only on the general form of
constraint algebras as opposed to specific corrections of individual
constraints are therefore reliable even in strong quantum regimes
where moments may be large. Signature change and possible consequences
presented in this paper are the main example.

\end{appendix}


\end{document}